\documentclass[ reprint, 
 aps,
 pra,
 showkeys,
 nofootinbib
]{revtex4-1}
\usepackage{amsfonts}
\usepackage{amssymb}
\usepackage{amsmath}
\usepackage{graphicx}
\usepackage{subfigure}
\usepackage{dcolumn}
\usepackage{bm}
\usepackage{setspace}
\usepackage[hang]{footmisc}
\usepackage[sort&compress]{natbib}
\usepackage{multirow}
\usepackage[sort&compress]{cleveref}
\usepackage{nomencl}
\usepackage[english]{babel}
\usepackage{xcolor, soul}
\usepackage{marginnote}

\sethlcolor{yellow}

\crefname{equation}{\unskip}{\unskip}
\crefname{figure}{\unskip}{\unskip}
\crefname{section}{\unskip}{\unskip}
\crefname{subsection}{\unskip}{\unskip}
\begin{document}
\let\ref\cref

\title{On the role of flexoelectricity in triboelectricity for randomly rough surfaces}

\author{B.N.J. Persson}
\affiliation{PGI-1, FZ J\"ulich, Germany}
\affiliation{www.MultiscaleConsulting.com, Wolfshovener str. 2, 52428 J\"ulich}

\begin{abstract}
I show how flexoelectricity result in a fluctuating surface electric potential when elastic solids with random
roughness are squeezed into contact. The flexoelectric potential may induce surface charge distributions
and hence contribute to triboelectricity. Using the developed theory I analyze the Kelvin Force Microscopy data
of Baytekin et al for the electric potential above a polydimethylsiloxane
(PDMS) surface after it was peeled away from another PDMS surface.
\end{abstract}

\maketitle
\makenomenclature


{\bf 1 Introduction}

When two solids in adhesive contact are pulled apart they are often
left charged. More generally, fluctuating charge distributions (both positive and negative charge distributions
on both solids) are left on the surfaces after sliding contact or pull-off\cite{Bay}. 
This triboelectric effect has been known for more than 25 centuries and has
important practical implications. However, the origin of the charge separation is not well understood\cite{a}.

In a recent paper Mizzi et al\cite{Mizzi} have proposed that triboelectricity often result from flexoelectricity.
The flexoelectric effect consist of the linear coupling between strain gradient and electric polarization\cite{p1,p2}.
When two solids are squeezed into contact very large strain gradients will occur in the asperity
contact regions, and Mizzi et al have shown that this may result in large flexoelectric potential differences at the
nanoscale. If free charges exist (in the solids or in the atmosphere)
they may move or rearrange in such a way as to screen the flexoelectric potential field,
and one may end up with charge distributions at, or close to, the solid surfaces which follow the spatial variation of the 
flexoelectric potential field.

Mizzi et al studied how the flexoelectric surface potential depends on the normal force
when a  sphere is squeezed against an elastic half-space. For this they 
used the well known results for the strain at Hertz (or JKR) sphere-flat type of contact\cite{Johnson}.
However, the contact between two randomly rough surfaces
cannot be described by a model assuming independent Hertzian contact regions since  
the long-ranged elastic coupling between the contact regions strongly affect the nature of the contact\cite{On,Mueser}. 
In this paper I show how the flexoelectric potential can be calculated for the contact
between elastic solids with randomly rough surfaces. 

\vskip 0.3cm
{\bf 2 Theory}

Mizzi et al (see Ref. \cite{Mizzi}) have shown that the normal component of the electric
field induced by a flexoelectric coupling in an isotropic nonpiezoelectric half-space 
is given (approximately) by
$$E_z = -f \left (\epsilon_{ii,z} +2 \epsilon_{zi,i}\right )\eqno(1)$$
where $\epsilon_{ij,k} = \partial \epsilon_{ij}/\partial x_k$, where 
$$\epsilon_{ij}={1\over 2} \left ({\partial u_i \over \partial x_j}+{\partial u_j \over \partial x_i}\right )\eqno(2)$$ 
is the strain. In (1) summation over repeated indicies is implicitly understood.
The flexocoupling voltage $f$ can be both positive or negative, and for polymers $|f|$ is typically in the range 
$5-300 \ {\rm V}$ (see Table S2 in Ref. \cite{Mizzi}). 
Substituting (2) in (1) gives
$$E_z = -f \left (\nabla^2 u_z +2 {\partial \over \partial z} \nabla \cdot {\bf u}\right )\eqno(3)$$
Assuming a solid with homogeneous and isotropic elastic properties,
from the theory of elasticity the displacement field ${\bf u}$ satisfies
$$\rho {\partial^2 {\bf u}\over \partial t^2} = \mu \nabla^2 {\bf u}+ (\mu+\lambda) \nabla \nabla \cdot  {\bf u}\eqno(4)$$
where $\rho$ is the mass density, and
where the Lame constants $\lambda$ and $\mu$ can be related to the Young's elastic modulus $E$ and the Poisson ratio $\nu$
via
$$\lambda = {\nu E \over (1+\nu)(1-2\nu)}, \ \ \ \ \ \mu = { E \over 2(1+\nu)}\eqno(5)$$
In principle the equation determining the deformation field ${\bf u}$ should be influenced
by the flexoelectric field, and the charge rearrangement which may
occur as result of it, but this effect will be assumed small in what follows.
We neglect the time dependency so that from (4): 
$$\nabla^2 {\bf u} = - \left (1+{\lambda\over \mu}\right )  \nabla \nabla \cdot  {\bf u}\eqno(6)$$
Using (3) and (6) gives
$$E_z = - \left ( 1-{\lambda\over \mu}\right )f {\partial \over \partial z}  \nabla \cdot {\bf u} \eqno(7)$$
The electric potential $\phi$ at the surface ($z=0$) relative to far inside the solid ($z=\infty$) is equal to
$$\phi=-\int_0^\infty dz \ E_z = - \left ( 1-{\lambda\over \mu}\right ) f \nabla \cdot {\bf u} \eqno(8)$$
where $\nabla \cdot {\bf u}$ is evaluated for $z=0$. 

For randomly rough surfaces the electric potential $\phi({\bf x})$ will vary in a complex way with the coordinate
${\bf x} = (x,y)$ on the surface $z=0$. Here we consider first the mean square of the (fluctuating) electric potential,
$\langle \phi^2({\bf x}) \rangle$, and then the electric potential power spectrum, which
contains more information about the fluctuating electric potential.

\vskip 0.1cm
{\bf 2.1 Mean square fluctuation of the electric potential}

Consider $\langle \phi^2  \rangle$, where $\langle .. \rangle$ 
stands for ensemble average. The ensemble average of $\phi^2({\bf x})$ is assumed to be independent of ${\bf x}$ so that
$$\langle \phi^2  \rangle = {1\over A_0} \int d^2 x \ \langle \phi^2 ({\bf x})\rangle , \eqno(9)$$
where $A_0$ is the surface area.
We write
$$\phi ({\bf x}) = \int d^2q \ \phi({\bf q}) e^{i{\bf q}\cdot {\bf x}} . \eqno(10)$$
From (9) and (10) we get
$$\langle \phi^2  \rangle = {(2\pi )^2\over A_0} \int d^2q \ \langle \phi(-{\bf q}) \phi({\bf q})\rangle \eqno(11)$$

We can calculate $\nabla \cdot {\bf u}$ using the formalism presented in Appendix A in Ref. \cite{BP}.
Assume that a stress $\sigma_i ({\bf x}, t)$ act on the surface of an elastic half space. We write
$$\sigma_i ({\bf x}, t) = \int d^2q d\omega \ \sigma_i ({\bf q}, \omega) e^{i({\bf q}\cdot {\bf x} - \omega t)}\eqno(12)$$
We note here that although we have assumed above that the time dependency of the flexoelectric field can be neglected,
it is in the present approach necessary to include it in the calculation of $\nabla \cdot {\bf u}$,
and only at the end let $\omega \rightarrow 0$ corresponding to a time-independent problem.

The displacement field in the solid for $z>0$ is written as [see Appendix A in Ref.  \cite{BP}]:
$${\bf u} = {\bf p}A+{\bf K}B +{\bf p}\times{\bf K} C\eqno(13)$$
where ${\bf p}=-i\nabla$ and where ${\bf K} = {\bf n}\times {\bf p}$, where ${\bf n}$ is a unit vector
normal to the surface pointing along the $z$-axis. Thus we get
$$\nabla \cdot {\bf u} = i {\bf p} \cdot {\bf u} = i p^2 A = -i \nabla^2 A\eqno(14)$$
The scalar potential $A$ satisfies the wave equation [see (A4) in Ref. \cite{BP}], which with the time variable
Fourier transformed, takes the form:
$$\nabla^2 A +{\omega^2 \over c_{\rm L}^2} A=0$$
Using this in (14) we get 
$$\nabla \cdot {\bf u} = i {\omega^2 \over c_{\rm L}^2} A\eqno(15)$$
Using (A19) in Ref. \cite{BP} we get for $z=0$, with the ${\bf x}$ and $t$ dependency Fourier transformed,
$$ i {\omega^2 \over c_{\rm L}^2} A=  {\omega^2 \over c_{\rm L}^2}  {1\over \mu S} \left [2 p_{\rm T} {\bf q}+
\left ({\omega^2 \over c_{\rm T}^2} - 2q^2\right ){\bf n}\right ] \cdot {\bf \sigma}\eqno(16)$$
where
$$S=\left ({\omega^2\over c_{\rm T}^2}-2q^2\right )^2+4q^2 p_{\rm T} p_{\rm L}\eqno(17)$$
where
$$p_T = \left ({\omega^2 \over c_{\rm T}^2}-q^2+i0^+\right )^{1/2}, \ \ \ \ \
p_L = \left ({\omega^2 \over c_{\rm L}^2}-q^2+i0^+\right )^{1/2}$$
where $c_{\rm T}$ and $c_{\rm L}$ are the transverse and longitudinal sound velocities, respectively.
Using (16) and (17) for $\omega \rightarrow 0$ we get
$$ i {\omega^2 \over c_{\rm L}^2} A= {1\over \lambda +\mu} \left (-i\hat q +{\bf n}\right )\cdot {\bf \sigma}\eqno(18)$$
where $\hat q = {\bf q}/q$. If we denote ${\bf e} = -i\hat q +{\bf n}$ we get from
(8), (11), (15) and (18):
$$\langle \phi^2  \rangle =   \left ({\kappa f \over E} \right )^2  {(2\pi )^2\over A_0} 
\int d^2q \ e^*_i e_j \langle \sigma_i (-{\bf q}) \sigma_j ({\bf q}) \rangle \eqno(19)$$
where
$$\kappa = {(\lambda-\mu) E \over \mu (\mu+\lambda)} = 2 (4\nu-1)(1+\nu)\eqno(20)$$
For rubber-like materials $\nu \approx 0.5$ giving $\kappa \approx 3$. 

We now consider the simplest case of pull-off without sliding. In this case the stress 
will be approximately normal to the surface and (19) reduces to
$$\langle \phi^2  \rangle =  \left ({\kappa f \over E }\right )^2  {(2\pi )^2\over A_0} 
\int d^2q \  \langle \sigma (-{\bf q}) \sigma ({\bf q}) \rangle \eqno(21)$$
where $\sigma  ({\bf q})$ now denote the normal stress component. 
Now, since
$$\int d^2q \  \langle \sigma (-{\bf q}) \sigma ({\bf q}) \rangle = {A_0\over (2 \pi )^2} \langle \sigma^2({\bf x})\rangle \eqno(22)$$
we can write
$$\langle \phi^2  \rangle =  \left ({\kappa f \over E }\right )^2  \langle \sigma^2 \rangle  \eqno(23)$$
If $P(\sigma ,p_0)$ is the probability distribution of stresses at the interface, which depends on the applied stress
$p_0$, then
$$\langle \sigma^2 \rangle = \int_{-\infty}^\infty d\sigma \ \sigma^2 P(\sigma ,p_0 ) \eqno(24)$$
The probability distribution $P(\sigma ,p_0 )$ can be calculated for randomly rough surfaces, both with and without
the adhesion, using the Persson contact mechanics theory\cite{rough}, or using numerically (exact) methods such as
the boundary element method\cite{Mueser,Ref8}. Here we consider first the non-adhesive contact between an elastic half-space
and a rigid countersurface, where the roughness is characterized by the surface roughness power spectrum $C(q)$.
For non-adhesive contact $P(\sigma ,p_0) =0$ for $\sigma < 0$ and for $\sigma > 0$ (see Ref. \cite{Yang}):
$$P = {1\over s (2\pi )^{1/2}} \left [e^{-(\sigma-p_0)^2/ (2 s^2)}-
e^{-(\sigma+p_0)^2/ (2s^2)} \right ],  \eqno(25)$$
where $s=E^* h' /2$, where $E^*=E/(1-\nu^2)$ and $h' = \langle (\nabla h)^2 \rangle^{1/2}$ is the rms-slope of the rough 
surface with the surface profile $z=h({\bf x})$.
For $p_0/E^* << 1$ the probability distribution reduces to
$$P \approx \left ( {2\over \pi}\right )^{1/2} {\sigma p_0 \over s^3} {\rm exp}\left (-{ \sigma^2 \over 2 s^2}\right ) \eqno(26)$$
In this limit we get from (24) and (26):
$$\langle \sigma^2 \rangle = 2 \left ({2 \over \pi} \right )^{1/2} s p_0\eqno(27)$$
so that
$$\langle \phi^2  \rangle =  \left ({\kappa f \over E }\right )^2 2 \left ({2 \over \pi} \right)^{1/2} s p_0\eqno(28)$$
or
$$\langle \phi^2  \rangle = \left [\left ({32\over \pi}\right )^{1/2} {1+\nu \over 1-\nu} (1-4\nu )^2\right ] f^2 h' {p_0 \over E}\eqno(29)$$ 

\begin{figure}
\includegraphics[width=0.5\columnwidth]{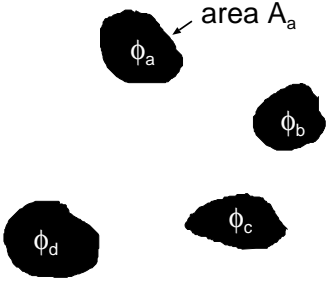}
\caption{\label{pic1.eps}
Contact patches between two elastic solids with random roughness.
The electric potential in the contact regions 
fluctuates between positive and negative values. 
The average contact patch area is denoted by $A_1$
and the rms electric potential in a contact patch by $\phi_1$. 
}
\end{figure}

In the limit when $A/A_0 << 1$, and when the surface roughness power spectrum has a wide roll-off region,
the contact regions will consist of a low concentration of small contact patches as indicated in Fig. \ref{pic1.eps}.
If the number of contact patches are denoted by $N$ and if the average area of a contact patch is $A_1$ then
$A=NA_1$. If the mean square value of the voltage at the surface of a contact region is $\phi_1^2$ then 
$$\langle \phi^2 \rangle A_0 = N  \phi_1^2 A_1$$ 
or 
$$\phi_1^2 = \langle \phi^2 \rangle {A_0\over A}\eqno(30)$$
Using that for $A/A_0 << 1$ we have\cite{rough}
$${A\over A_0} \approx {2 p_0 \over h' E^*}\eqno(31)$$
From (29)-(31) we get
$$\phi_1 \approx \xi |f| h' \eqno(32)$$ 
where 
$$\xi= \left ({8\over \pi}\right )^{1/4} {|1-4\nu|\over 1-\nu} \eqno(33)$$

Another interesting limiting case is contact with adhesion when the adhesion is so strong
as to pull the solids into complete contact. This limit is easy to study: First note that complete
contact prevail as $p_0 \rightarrow \infty$ and in this limit we can neglect the second term
in (25). This gives a Gaussian-like probability distribution centered at $\sigma = p_0$.
However, with adhesion we consider the case without an applied pressure i.e., $p_0=0$. Thus for
adhesion and assuming complete contact with $p_0=0$ we have the (exact) 
stress probability distribution
$$P = {1\over s (2\pi )^{1/2}} e^{-\sigma^2/ (2 s^2)}\eqno(34)$$
and hence
$$\langle \sigma^2  \rangle = {1\over s (2\pi )^{1/2}}   
\int_{-\infty}^\infty d\sigma \ \sigma^2 e^{-\sigma^2/ (2 s^2)} = s^2 \eqno(35)$$ 
Substituting (35) in (23) gives
$$\langle \phi^2  \rangle =   \left ({\kappa f h' \over 2(1-\nu^2) }\right )^2 \eqno(36)$$
or 
$$\langle \phi^2  \rangle^{1/2} = \xi' |f| h'\eqno(37)$$ 
where
$$\xi'=  {4\nu-1\over 1-\nu}\eqno(38)$$

For rubber-like materials $\nu \approx 0.5$ giving $\xi \approx 2.5$ and $\xi' \approx 2.0$.
As an example, for natural rubber $|f| \approx 20 \ {\rm V}$ and if we assume 
the rms slope $h' \approx 0.1$ we get the average electric potential 
in the asperity contact regions $\phi_1 \approx 5 \ {\rm V}$, and for complete contact
the rms surface electric potential $\langle \phi^2  \rangle^{1/2} \approx 4  \ {\rm V}$.  

\vskip 0.1cm
{\bf 2.2 Electric potential power spectrum}

Let us now study the
electric potential power spectrum
$$C_{\phi \phi} ({\bf q}) = {(2 \pi)^2\over A_0} \langle |\phi ({\bf q})|^2\rangle  \eqno(39)$$
We can calculate $C_{\phi \phi}$ from the theory presented above. Thus from (8), (15) and (18):
$$\phi({\bf q}) =  - \left ( 1-{\lambda\over \mu}\right ) f {1\over \lambda +\mu} {\bf e} \cdot {\bf \sigma} \eqno(40)$$
Thus we get
$$C_{\phi \phi} ({\bf q}) = {(2 \pi)^2\over A_0} \left ( {\mu-\lambda\over \mu (\mu+\lambda)}\right )^2 
f^2 e_i^* e_j \langle \sigma_i(-{\bf q}) \sigma_j({\bf q})\rangle \eqno(41)$$
If we assume no shear forces we get
$$C_{\phi \phi} ({\bf q}) = {(2 \pi)^2\over A_0} \left ( {\mu-\lambda\over \mu (\mu+\lambda)}\right )^2 
f^2 \langle \sigma(-{\bf q}) \sigma({\bf q})\rangle \eqno(42)$$
where $\sigma ({\bf q})$ is the normal stress. In Ref. \cite{On} we have shown that to a good approximation
$$\langle \sigma(-{\bf q}) \sigma({\bf q})\rangle \approx 
\left ({\mu \over 1-\nu}\right )^2  {A_0\over (2 \pi)^2} q^2 C(q) P(q) \eqno(43)$$
Here $C(q)$ is the surface roughness power spectrum, and $P(q) = A(q)/A_0$ is the
relative contact area when only the roughness components with the wavenumber smaller than $q$
is included when calculating $A(q)$ (see Ref. \cite{rough}).  

Substituting (43) in (42) gives
$$C_{\phi \phi}  \approx \left ({\mu -\lambda \over \mu+\lambda}\right )^2  
\left ({f\over 1-\nu} \right )^2 q^2 C(q) P(q)\eqno(44)$$
For rubber materials $\nu \approx 0.5$ and $\lambda \approx \infty$ giving
$$C_{\phi \phi}  \approx  4 f^2 q^2 C(q) P(q)\eqno(45)$$
If we assume complete contact between the solids at the interface then $P(q)=1$ and
$$C_{\phi \phi}  \approx  4 f^2 q^2 C(q) \eqno(46)$$

Many surfaces are self affine fractal with a roll-off region for $q<q_0$. For such surfaces\cite{fractal}:
$$C=C_0  \ \ \ \ {\rm for}  \ \ \ \  q<q_0 \eqno(47)$$
$$C=C_0 \left ({q\over q_0}\right )^{-2(1+H)}  \ \ \ \  {\rm for}  \ \ \ \  q>q_0\eqno(48)$$
where
$$C_0 = {1\over \pi} {H\over 1+H} {\langle h^2\rangle \over q_0^2}\eqno(49)$$
where $\langle h^2 \rangle$ is the mean square roughness and where
the Hurst exponent $H$ is between 0 and 1 but typically $H\approx 1$ (see Ref. \cite{fractal}).
The latter is expected for sandblasted surfaces and, at least in some cases, 
for contaminated surfaces\cite{conta},
which can be considered as generated by a process opposite to sandblasting
(depositing of particles rather than removal of particles).  
Substituting (47) and (48) in (46)
gives an electric potential power spectrum which scales $\sim q^{-2H}$ for $q>q_0$
and $q^2$ for $q<q_0$. For $q=q_0$
$$C_{\phi \phi} \approx  4 f^2 q_0^2 C_0 \eqno(50)$$
which gives
$$\langle h^2 \rangle = {\pi \over 4} {1+H\over H} f^{-2} C_{\phi \phi}\eqno(51)$$
where $C_{\phi \phi}$ is evaluated for $q=q_0$.

\begin{figure}
\includegraphics[width=0.9\columnwidth]{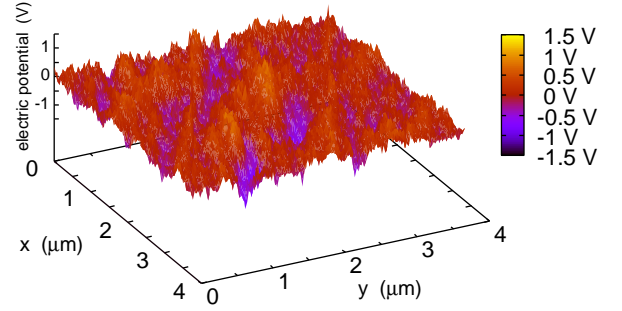}
\caption{\label{topo.eps}
The measured electric potential a distance $\approx 100 \ {\rm nm}$ above a
PDMS surface\cite{Bay}.
}
\end{figure}

\begin{figure}
\includegraphics[width=0.9\columnwidth]{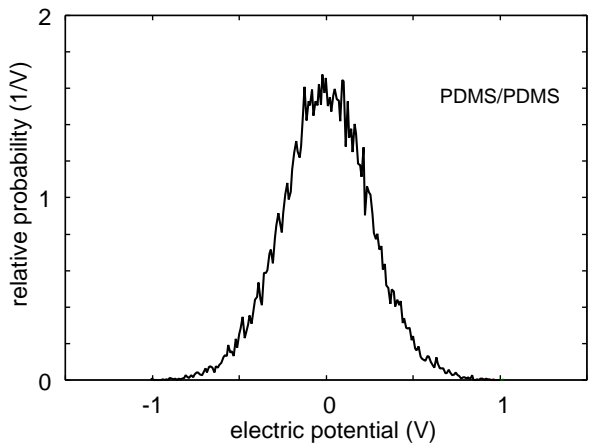}
\caption{\label{1Voltage.2PVoltage.eps}
The probability distribution of the electric potential $P(\phi)$ a distance $d\approx 100 \ {\rm nm}$
above a PDMS surface.
The rms voltage fluctuation from the measured data is $\langle \phi^2 \rangle^{1/2} \approx 0.26 \ {\rm V}$.
}
\end{figure}

\begin{figure}
\includegraphics[width=0.9\columnwidth]{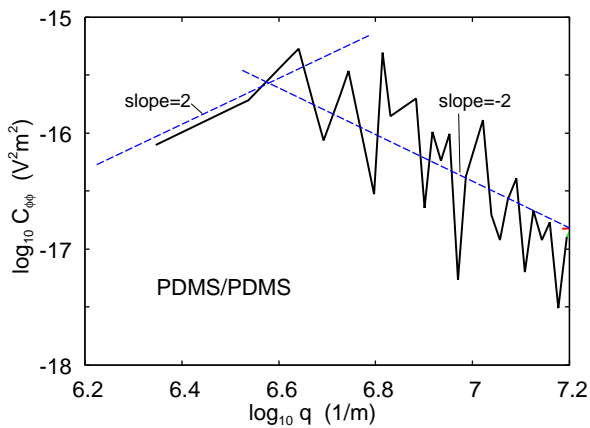}
\caption{\label{1logq.2logCVV.eps}
The electric potential power spectrum $C_{\phi \phi}$ as a function of the wavenumber $q$
(log-log scale). The blue dashed lines have the slope $+2$ and $-2$.
}
\end{figure}

\vskip 0.3cm
{\bf 3 Discussion}

The flexoelectric effect gives rise to an electric field in the surface region between contacting solids. 
This field will drive charges (either from the gas phase or from the solids) 
to rearrange in such a way as to screen out the electric field as completely as possible. 
Following Mizzi et al we will assume that when the solid bodies are separated this will result in a 
surface distribution of charges which will generate an electric potential outside of the solids.
The charge distribution will of course decay with increasing time but a short time after the
surface separation it may result in an electric potential distribution of similar form as that
generated by the flexoelectric effect. This electric potential can be studied using Kelvin Force Microscopy.

In Ref. \cite{Bay} Baytekin et al have used Kelvin Force Microscopy in a study of the electric potential 
$\phi ({\bf x})$ at a distance $\approx 100 \ {\rm nm}$ above
a PDMS surface, a short time after separating it from the contact with another PDMS surface.
The PDMS surfaces was prepared by molding the rubber against an atomically flat (silanized) silicone wafer. However,
the silicone wafer was exposed to the normal atmosphere and may have a contamination film so the
PDMS sheets may have surface roughness, with a rms roughness amplitude of order a few nanometer. 

Because of the small surface roughness the two PDMS sheets are likely to be in complete adhesive contact.
After separation the surfaces have charge distributions which oscillates between positive and negative values
(see Fig. \ref{topo.eps}) such that the net charge is small compared to the total number of charges. This is clear from
the probability distribution $P(\phi)$ of the electric potential shown in Fig. \ref{1Voltage.2PVoltage.eps}. Note that
$P(\phi)$ is nearly a perfect Gaussian centered at $\phi\approx 0$ i.e. the net charge is very small.

Fig. \ref{1logq.2logCVV.eps} shows the electric potential power spectrum. The data is very noisy due to the rather small
number of Kelvin Force Microscopy data points ($152 \times 152$). We do not show results 
for $q>1.6\times 10^7 \ {\rm m}^{-1}$ (or ${\rm log}_{10} q > 7.2$) because $C_{\phi \phi}$ for large $q$ is influenced by the fact that 
the scanning tip was located $\approx 100 \ {\rm nm}$ above the PDMS surface. Note that
for small wavenumber $q < q_0 \approx 4\times 10^6 \ {\rm m}^{-1}$ the power spectrum increases like $q^2$ with the wavenumber, while for
$q>q_0$ it decreases roughly as $q^{-2}$; both results are expected from the theory above if the Hurst exponent $H \approx 1$
and if a flat roll-off region occur in the roughness power spectrum for $q<q_0$. Baytekin et al also performed studies
for PDMS pulled off from a smooth polycarbonate (PC) surface, but the results are very similar as for PDMS against PDMS.  
We do note however that the region for $q<q_0$ in Fig. \ref{1logq.2logCVV.eps} is very uncertain due to the small number
of long-wavelength roughness components. Thus, a more accurate study require Kelvin Force Microscopy measurements over a larger surface area.

We can use (51) to estimate the rms-roughness necessary in order to reproduce the magnitude of the observed electric potential
power spectrum. Thus for $q=q_0$ from Fig. \ref{1logq.2logCVV.eps} we get $C_{\phi \phi} \approx 2\times10^{-16} \ {\rm V^2 m^2}$
and using (51) with $H=1$ this gives $\langle h^2 \rangle^{1/2} \approx 6 \ {\rm nm}$ if $f=3 \ {\rm V}$.
This values is very reasonable for a surface where the roughness is produced by
a contamination film due to the exposure of the wafer to the normal atmosphere. 
The roll-off wavelength $\lambda_0 = 2\pi /q_0 \approx 1 \ {\rm \mu m}$ also appear very reasonable.

\vskip 0.3cm
{\bf 4 Summary and  conclusion}

We have presented a theory for the electric potential at a surface produced by flexoelectricity for elastic solids
with randomly rough surfaces. We have calculated the power spectrum of the electric potential $\phi({\bf x})$.
In the light of the theory we have discussed the experimental contact electrification results of Baytekin et al and found good 
correlation with the theory predictions. 

\vskip 0.3cm
{\bf Acknowledgments}

I thank E. Tosatti for drawing my attention to the paper of C.A. Mizzi et al.
I thank C.A. Mizzi and L.D. Marks for useful communication.
I thank B. Baytekin, H.T. Baytekin and  B.A. Grzybowski for kindly supplying the KFM potential maps used in calculating
the results shown in Fig. 2-4.

\end{document}